\documentclass[12pt]{article}

\usepackage{times}
\usepackage{graphicx}

\newenvironment{sciabstract}{%
\begin{quote} \bf}
{\end{quote}}

\newcounter{lastnote}

\title{X-Ray Flares from Postmerger Millisecond Pulsars}

\author
{Z. G. Dai,$^{1\ast}$ X. Y. Wang$^{1}$, X. F. Wu$^{2}$, B. Zhang$^{3}$\\
\\
\normalsize{$^{1}$Department of Astronomy, Nanjing
University, Nanjing 210093, China}\\
\normalsize{$^{2}$Purple Mountain Observatory, Chinese Academy
of Sciences, Nanjing 210008, China;}\\
\normalsize{J-CPNPC of PMO-NJU, Chinese Academy
of Sciences, Nanjing 210008, China}\\
\normalsize{$^{3}$Department of Physics, University of Nevada,
Las Vegas, NV 89154, USA}\\
\\
\normalsize{$^\ast$ To whom correspondence should be addressed. E-mail:
dzg@nju.edu.cn} }

\date{}

\begin{document}


\baselineskip24pt


\maketitle


\begin{sciabstract}
Recent observations support the suggestion that short-duration gamma-ray bursts
are produced by compact star mergers. The X-ray flares discovered in two short
gamma-ray bursts last much longer than the previously proposed postmerger
energy release time scales. Here we show that they can be produced by
differentially rotating, millisecond pulsars after the mergers of binary
neutron stars. The differential rotation leads to windup of interior poloidal
magnetic fields and the resulting toroidal fields are strong enough to float up
and break through the stellar surface. Magnetic reconnection--driven explosive
events then occur, leading to multiple X-ray flares minutes after the original
gamma-ray burst.
\end{sciabstract}

Gamma-ray bursts (GRBs) are flashes of gamma rays occurring at the cosmological
distances. They fall into two classes ({\it 1}): short-duration ($<2$ s)
hard-spectrum bursts and long-duration soft-spectrum bursts. Long GRBs result
from core collapses of massive stars ({\it 2}), and short GRBs appear to be
produced in mergers of neutron star binaries or black hole-neutron star
binaries ({\it 3-9}). Recently thanks to accurate localizations of several
short GRBs ({\it 3,6,8}) by {\em Swift} and {\em High Energy Transient
Explorer-2} ({\em HETE-2}), the multi-wavelength afterglows from these events
have been detected and the associated host galaxies have been identified. The
observations provide a few pieces of evidence in favor of the binary compact
object merger origin of short GRBs ({\it 10-12}). Because it takes $\sim 0.1-1$
billions years of gravitational wave radiation before the binary coalesces, at
least some short GRB host galaxies should contain a relatively old stellar
population. Because neutron stars in the binary system usually receive a very
high natal velocity, the merger site is preferably at the outskirt of the host
galaxy, and the circumburst medium density is likely low. These characteristics
have been revealed by recent observations: First, the identified elliptical
galaxies associated with GRB 050509B ({\it 3,4}) and GRB 050724 ({\it 8,9})
suggest that these hosts are early type galaxies with a low star-formation
rate, ruling out progenitor models invoking active star formation. Second, the
nondetection of any supernova signal from GRB 050709 indicates that short
bursts are not associated with collapses of massive stars ({\it 5,7}).  Third,
afterglow modeling of GRB 050709 suggests a low density environment ({\it 13}),
which is consistent with that of the outskirt of the host galaxy or that of an
intergalactic medium.

However, the above merger origin was recently challenged by the discovery of
X-ray flares occurring after two short bursts. X-ray flares were discovered to
occur at least $\sim 100\,$s after the triggers of the short GRB 050709 ({\it
5}) and GRB 050724 ({\it 8}). These flares require that the central engine is
in long-lasting activity. This requirement conflicts with the current models
involving neutron star-neutron star mergers ({\it 14,15}) or neutron star-black
hole mergers ({\it 16}), because all these models are attached to a common
postmerger picture that invokes a black hole surrounded by a torus. The
predicted typical time scales for energy release are much shorter than $\ge
100$ s as observed in GRBs 050709 and 050724. Therefore, understanding the
origin of X-ray flares from short bursts is currently of great interest. Here
we show that such flares can be produced by differentially rotating,
millisecond pulsars with typical surface magnetic fields that occur after the
mergers of binary neutron stars.

In the conventional scenarios of short bursts ({\it 10-12}), after the merger
of a neutron star binary, a stellar-mass black hole is formed with a transient
torus of mass $\sim 1-10\%$ of the total. These scenarios are valid if the
total mass ($\sim 2.5-2.8M_\odot$, where $M_\odot$ is the solar mass) of the
postmerger object is larger than the maximum mass of a nonrotating
Tolman-Oppenheimer-Volkoff neutron star, $M_{{\max},0}$. This is valid if the
nuclear equation of state (EOS) is soft to moderately stiff ({\it 17}).
However, the total mass of the postmerger object is smaller than $M_{{\max},0}$
for very stiff EOSs (e.g., as predicted by mean field theory) ({\it 17}).
Timing observations of the millisecond pulsar J0751+1807 in a circular binary
system with a helium white-dwarf companion ({\it 18}) reveal the existence of a
neutron star with mass of $2.1\pm 0.2M_\odot$ (at the $1\sigma$ confidence
level). This measurement implies that the maximum mass of nonrotating neutron
stars must be larger than $2.1M_\odot$ so that stiff EOSs are favored.
Furthermore, recent general relativistic numerical simulations ({\it 17,19})
have shown that for stiff to very stiff nuclear EOSs, the postmerger object is
indeed a differentially rotating massive neutron star with period of $\sim 1$
ms, because uniform rotation and differential rotation can support a maximum
mass $\sim 20\%$ and $\sim 50\%$ higher than $M_{{\max},0}$, respectively. It
is therefore reasonable to assume the existence of a differentially rotating
millisecond pulsar after a double neutron star merger. Such a pulsar should
also be surrounded by a hot torus with mass $\sim 0.01-0.1M_\odot$. Similar to
the previous scenarios, a short burst may be produced by the Parker instability
in the torus ({\it 11}) or the annihilation of neutrinos emitted from the torus
({\it 12}).

After the GRB trigger, differential rotation starts to wind the interior
magnetic field into a toroidal field ({\it 20,21}). To represent physical
processes of windup and floating of the magnetic field, we consider a simple
two-component model in which the star is divided into two zones with a boundary
at the radius $R_c\simeq 0.5R_*$ (where $R_*$ is the stellar radius): the core
and the shell components. Their moments of inertia are $I_c$ and $I_s$ and
their angular (rotation) velocities are $\Omega_c$ and $\Omega_s$,
respectively. The differential angular velocity is then
$\Delta\Omega=\Omega_c-\Omega_s$ and its initial value (marked by a subscript
zero) is taken as $(\Delta\Omega)_0=A_0\Omega_{s,0}$ (where $A_0$ is the ratio
of the initial differential angular velocity to the shell's initial angular
velocity). If the radial magnetic field component is $B_r$, then the toroidal
field component $B_\phi$ increases as
\begin{equation}
\frac{dB_\phi}{dt}=(\Delta\Omega)B_r.
\end{equation}
There is a magnetic torque , $T_m=(2/3)R_c^3B_rB_\phi$, acting between the core
and shell ({\it 22}). This torque opposes the differential rotation. Another
torque results from magnetic dipole radiation,
$T_d=2B_s^2R_*^6\Omega_s^3/(3c^3)$, where $c$ is the speed of light and
$B_s=\epsilon B_r$ (here $\epsilon$ is defined by the ratio of the effective
surface dipole field strength to the radial field strength). Under action of
these two torques, the angular velocities of the shell and the core components
evolve according to
\begin{equation}
I_s\frac{d\Omega_s}{dt}=T_m-T_d
\end{equation}
and
\begin{equation}
I_c\frac{d\Omega_c}{dt}=-T_m,
\end{equation}
respectively. The torque from magnetic dipole radiation can be neglected if
$B_\phi\gg B_r\epsilon^2(R_*/R_c)^3\\ \times (R_*\Omega_s/c)^3$. This condition
is easily satisfied at the time $t\gg t_0\equiv
\epsilon^2A_0^{-1}\Omega_{s,0}^{-1}(R_*/R_c)^3$ (where $t_0$ is $\sim$0.2 ms
for typical parameters). Thus, from equations (1-3), we obtained
\begin{equation}
\frac{d^2\Delta\Omega}{dt^2}=-\frac{2I}{3I_cI_s}R_c^3B_r^2(\Delta\Omega),
\end{equation}
where $I=I_c+I_s$ is the total moment of inertia of the star.
Letting
\begin{equation}
\tau = \left(\frac{2I}{3I_cI_s}R_c^3B_r^2\right)^{-1/2}\simeq
2.3\times 10^5(\epsilon/0.3)B_{s,8}^{-1}\,{\rm s},
\end{equation}
where $I_c\simeq I_s=10^{45}\,{\rm g}\,{\rm cm}^2$ and $R_*=10^6$ cm are taken
and $B_{s,8}$ is in units of $10^8$ G, we found a solution of equation (4),
\begin{equation}
\Delta\Omega=A_0\Omega_{s,0}\cos (t/\tau).
\end{equation}
This indicates that differential rotation would behave as a
resonator if there is no energy dissipation.

The increasing toroidal field becomes unstable because of the buoyancy
effect when $B_\phi=B_b\simeq 10^{17}\,$G ({\it 20}). This corresponds
to the time,
\begin{equation}
t_b = \frac{B_b}{B_rA_0\Omega_{s,0}}\simeq 4.8\times
10^4(\epsilon/0.3)B_{s,8}^{-1}A_0^{-1}P_{s,0,{\rm ms}}\,{\rm s},
\end{equation}
where $P_{s,0,{\rm ms}}$ is the initial spin period of the shell component in
units of milliseconds. Comparing equations (5) and (7), we see that $t_b$ is
substantially less than $\tau$ for $A_0^{-1}P_{s,0,{\rm ms}}\le 1$ and that the
differential angular velocity $\Delta\Omega$ is approximately constant until
the time $t_b$. At this time, the buoyant force is just equal to the force from
antibuoyant stratification existing in the star. As the time increases, the
buoyant force acting on the toroid would begin to exceed the antibuoyant force
and the toroid will float up toward the stellar surface. The net force density
acting on this toroid is given by
\begin{equation}
f_b=\frac{B_rB_bA_0\Omega_{s,0}(t-t_b)}{4\pi c_s^2}g,
\end{equation}
where $c_s$ is the speed of sound of the embedding medium and $g$ is the
surface gravity. In terms of equation (8) and Newton's second law, we obtained
the buoyancy timescale for the toroid to float up and penetrate through the
stellar surface:
\begin{equation}
\Delta t_b = \left[\frac{12\pi \rho
R_*c_s^2}{B_rB_bgA_0\Omega_{s,0}}\right]^{1/3} \simeq
0.26(\epsilon/0.3)^{1/3}B_{s,8}^{-1/3}A_0^{-1/3}P_{s,0,{\rm ms}}^{1/3}\,{\rm
s},
\end{equation}
where $\rho\simeq 10^{14}\,{\rm g}\,{\rm cm}^{-3}$ is the mass
density of the embedding medium, and the typical values of the
speed of sound and the surface gravity are $10^{10}\,{\rm
cm}\,{\rm s}^{-1}$ and $10^{14}\,{\rm cm}\,{\rm s}^{-2}$,
respectively. This timescale is much shorter than $t_b$, suggesting that
the toroid, after its field strength reaches $B_b$, would rapidly
float up to the stellar surface.

Once penetrating through the surface, the toroidal fields with different
polarity may reconnect ({\it 20}), giving rise to an explosive event. Its
energy is
\begin{equation}
E_b=\frac{B_b^2}{8\pi}V_b\simeq 1.6\times 10^{51}\, {\rm
ergs}\left(\frac{V_b}{V_*}\right),
\end{equation}
where $V_b$ and $V_*$ are the toroid's volume and stellar volume, respectively.
This energy depends on the toroid's volume rather than on the initial magnetic
field and the stellar spin period. An upper limit to the outflow mass ejected
is estimated by
\begin{equation}
M_{b,{\rm max}} = f_bV_b/g \simeq 0.9\times 10^{-7}M_\odot
(\epsilon/0.3)^{-2/3} B_{s,8}^{2/3}A_0^{2/3}P_{s,0,{\rm
ms}}^{-2/3}\left(\frac{V_b}{V_*}\right).
\end{equation}
Because of an initial huge optical depth, the outflow will expand
relativistically and its minimum average Lorentz factor is
\begin{equation}
\Gamma_{b,{\rm min}}\simeq 1.0\times
10^4(\epsilon/0.3)^{2/3}B_{s,8}^{-2/3} A_0^{-2/3}P_{s,0,{\rm
ms}}^{2/3}.
\end{equation}

The X-ray flares observed at $t_{\rm flare}\sim t_b\simeq 100$ s after GRBs
050709 and 050724 require that the surface magnetic field of a central pulsar
$B_s \sim 4.8\times 10^{10} (\epsilon/0.3)A_0^{-1}P_{s,0,{\rm ms}}(t_{\rm
flare}/100\,{\rm s})^{-1}\,{\rm G}$. For typical values ({\it 19,22}) of the
model parameters (i.e., $A_0\sim 1$, $P_{s,0}\sim 1$ ms, and $\epsilon\sim
0.3$), this field strength is in the range of the surface magnetic fields of
isolated pulsars. Furthermore, it is characteristic of the stellar magnetic
field that has decayed in $\sim 0.1-1$ billion years before the merger of a
neutron star binary ({\it 23}). Inserting this field into equation (12), we
found the minimum average Lorentz factor of the outflow from a
magnetic-reconnection-driven explosion, $\Gamma_{b,{\rm min}}\sim 160(t_{\rm
flare}/100\,{\rm s})^{2/3}$, showing that the outflow is ultrarelativistic.
After the end of this event, a similar windup of the interior magnetic field
with $B_r$ would start again following the same processes described above,
leading to another explosion.

Collisions among the outflows with different Lorentz factors would produce late
internal shocks and X-ray flares ({\it 24,25}). These shocks must produce lower
energy photons than did the earlier internal shocks during the prompt GRB
phase. For the internal shock model, the characteristic synchrotron frequency
is $\nu_m\propto L^{1/2} R_{\rm sh}^{-1}\propto L^{1/2} \Gamma_b^{-2}\delta
t^{-1}$ (where $L$ is the luminosity, $R_{\rm sh}$ is the shock radius,
$\Gamma_b$ is the bulk Lorentz factor, and $\delta t$ is the time interval
between two adjacent energy shells that the central engine ejects). The late,
soft flare is the result of the combination of a lower luminosity and a longer
time interval (than that of the prompt emission, where $\delta t\sim t_b$ in
our flare model). In addition, as the stellar differential rotation weakens
(i.e., $A_0$ decreases), the time interval $\delta t$ and the outflow's Lorentz
factor $\Gamma_b$ increase (see equations 7 and 12). Because the maximum flux
density of the synchrotron radiation scales as $F_{\nu,{\rm max}}\propto
\Gamma_b^{-3}$ ({\it 24}), the flux density at frequency $\nu$ is
$F_\nu=F_{\nu,{\rm max}}(\nu/\nu_m)^{-(p-1)/2}\propto \Gamma_b^{-(2+p)}\delta
t^{-(p-1)/2}$ for $\nu>\nu_m$ in the slow-cooling case (where $p$ is the
spectral index of the shock-accelerated electrons) ({\it 26}). Thus, the flare
occurring at later times has a smaller flux density because of the larger
Lorentz factor and longer time interval. This result is consistent with the
observed reduced flaring activity of GRB 050724. Therefore, our model can
provide a self-consistent explanation for all the observations including the
energetics (see equation 10) and the temporal and spectral properties of the
X-ray flares.

Generally speaking, the surface magnetic field of the postmerger pulsar could
have a wider range than the preferred value invoked here to interpret the $\sim
100$ s flares in GRBs 050709 and 050724. For stronger fields, this would give
rise to multi-peaks in the prompt phase (as observed in some short GRBs) or, if
the flares are not bright enough, they may be masked by the steep decay
component of the prompt emission tail ({\it 25}). For weaker fields, the
putative flares occur much later and are energetically insignificant. This
would give rise to smoother X-ray afterglow lightcurves as observed in several
GRBs observed by {\em Swift} (e.g. GRB 050509B) ({\it 3}).

X-ray flares were observed in nearly a half of long {\em Swift} bursts ({\it
27,28}). Even though the two classes of bursts have different progenitors
(namely collapsars for long bursts and binary neutron star mergers for short
bursts), similar temporal properties (e.g. peak times and temporal indices
before and after the peaks) suggest that the X-ray flares may have a common
origin. Therefore, we suggest that some long bursts may originate from
moderately-magnetized millisecond pulsars with hyperaccreting accretion disks
after the collapses of massive stars and their X-ray flares are the result of
strong interior differential rotation of these pulsars. The differences in
duration, energetics and spectrum for the two classes of bursts would be due to
different accretion disks, e.g., a transient torus for short bursts ({\it
10-12}) and a fall-back accretion disk for long bursts ({\it 29,30}). When the
surface magnetic fields are strong enough, the spin down of this central engine
pulsar would provide energy injection to the postburst relativistic outflow
({\it 31}), which could interpret the late X-ray humps detected in many GRBs
({\it 25,28}).

\vspace{5mm}
{\bf References and Notes}
\begin{enumerate}

\item  C. Kouveliotou {\it et al.}, {\it Astrophys. J.} {\bf 413}, L101 (1993).

\item  For a review, see B. Zhang, P. M\'esz\'aros, {\it Int. J. Mod. Phys.} {\bf A19},
   2385 (2004).

\item  N. Gehrels {\it et al.}, {\it Nature} {\bf 437}, 851 (2005).

\item  J. S. Bloom  {\it et al.}, {\it Astrophys. J.} {\bf 638}, 354 (2006).

\item   D. B. Fox {\it et al.}, {\it Nature} {\bf 437}, 845 (2005).

\item  J. S. Villasenor {\it et al.}, {\it Nature} {\bf 437}, 855 (2005).

\item  J. Hjorth {\it et al.}, {\it Nature} {\bf 437}, 859 (2005).

\item  S. D. Barthelmy {\it et al.},  {\it Nature} {\bf 438}, 994 (2005).

\item  E. Berger {\it et al.}, {\it Nature} {\bf 438}, 988 (2005).

\item  D. Eichler, M. Livio, T. Piran, D. N. Schramm, {\it Nature} {\bf 340},
    126 (1989).

\item  R. Narayan, B. Paczy\'nski, T. Piran, {\it Astrophys. J.} {\bf 395}, L83
    (1992).

\item  R. Mochkovitch,  M. Hernanz, J. Isern,  X. Martin,
    {\it Nature} {\bf 361}, 236 (1993).

\item  A. Panaitescu,  {\it Mon. Not. R. Astron. Soc.}, submitted (preprint available at
    http://arXiv.org/astro-ph/0511588).

\item  S. Rosswog, E. Ramirez-Ruiz,  M. B. Davies,
    {\it Mon. Not. R. Astron. Soc.} {\bf 345}, 1077 (2003).

\item  M. A. Aloy, H.-T. Janka, E. M\"uller, {\it Astron. Astrophys.}
    {\bf 436}, 273 (2005).

\item  M. B. Davies, A. Levan, A. King, {\it Mon. Not. R. Astron. Soc.}
    {\bf 356}, 54 (2005).

\item  The soft EOS at high densities models the
    interaction of nucleons with a Reid soft-core potential, the
    moderately stiff EOS uses the two-body and three-body interactions,
    and the very stiff EOS models the nucleon interaction in terms of a mean
    scalar field. The effects of these EOSs, uniform rotation, and differential
    rotation on the maximum mass of neutron stars have been explored in ({\it 32}).

\item  D. J. Nice {\it et al.}, {\it Astrophys. J.} {\bf 634}, 1242 (2005).

\item  M. Shibata, K. Taniguchi, K. Ury${\rm \bar{u}}$, {\it Phys. Rev. D}
    {\bf 71}, 084021 (2005).

\item  W. Klu\'zniak, M.  Ruderman, {\it Astrophys. J.} {\bf 505}, L113 (1998).

\item  Z. G. Dai, T. Lu, {\it Phys. Rev. Lett.} {\bf 81}, 4301 (1998).

\item  H. C. Spruit, {\it Astron. Astrophys.} {\bf 341}, L1 (1999).

\item  P. Goldreich, A. Reisenegger, {\it Astrophys. J.} {\bf 395}, 250 (1992).

\item  Y. Z. Fan, D. M.  Wei, {\it Mon. Not. R. Astron. Soc.} {\bf 364}, L42 (2005).

\item  B. Zhang {\it et al.},  {\it Astrophys. J.}, in press (preprint available at
    http://arXiv.org/astro-ph/0508321).

\item  To calculate the X-ray flux, two spectral break frequencies should be
    considered ({\it 2}): the characteristic frequency $\nu_m$ and the cooling
    frequency $\nu_c$. For typical parameters in our model, $\nu_m<\nu_c$, implying
    the slow-cooling case.

\item  D. N. Burrows {\it et al.}, {\it Science} {\bf 309}, 1833 (2005).

\item  P. T. O'Brien {\it et al.}, {\it Astrophys. J.}, submitted (preprint available at
    http://arXiv.org/astro-ph/0601125).

\item  R. Popham, S. E. Woosley, C. Fryer, {\it Astrophys. J.} {\bf 518}, 356 (1999).

\item  A. I. MacFadyen, S. E. Woosley, {\it Astrophys. J.} {\bf 524}, 262 (1999).

\item  Z. G. Dai, T. Lu, {\it Astron. Astrophys.} {\bf 333}, L87 (1998).

\item  I. A. Morrison, T. W. Baumgarte, S. L. Shapiro, {\it Astrophys. J.} {\bf 610},
    941 (2004).

\item  We thank P. F. Chen and M. D. Ding for helpful discussions about solar
    flare models that have motivated us to consider X-ray flare mechanisms and
    Y. F. Huang for valuable suggestions. This work is supported by the National
    Natural Science Foundation of China (grant numbers: 10221001, 10233010, 10403002,
    and 10503012). B. Z. is supported by NASA (grant numbers: NNG05GB67G, NNG05GH92G,
    and NNG05GH91G).

\end{enumerate}

\end{document}